\documentclass[journal=apchd5,manuscript=article]{achemso}

\SectionNumbersOff

\usepackage[version=3]{mhchem} 

\usepackage[T1]{fontenc}       
\usepackage{braket} 

\author{Juan Pablo Vasco}
\affiliation{Department of Physics, Engineering Physics and Astronomy, Queen's University, Kingston, Ontario, Canada, K7L 3N6}
\email{jpvc@queensu.ca}
\author{Stephen Hughes}
\affiliation{Department of Physics, Engineering Physics and Astronomy, Queen's University, Kingston, Ontario, Canada, K7L 3N6}
\email{shughes@queensu.ca}

\title{Anderson localization in disordered \textit{LN} photonic crystal slab cavities}


\begin{document}


\begin{abstract}
We present a detailed theoretical study of the effects of structural disorder on \textit{LN} photonic crystal slab cavities, ranging from short to long length  scales (\textit{N}=$3$-$35$ cavity lengths), using a fully three-dimensional Bloch mode expansion technique. We compute the optical density of states (DOS), quality factors and effective mode volumes of the cavity modes, with and without disorder, and compare with the localized  modes of the corresponding disordered photonic crystal waveguide. We demonstrate how the  quality factors and effective mode volumes saturate at a specific cavity length and become bounded by the corresponding values of the Anderson modes appearing in the disordered waveguide. By means of the intensity fluctuation criterion, we observe Anderson-like localization for cavity lengths  larger than around \textit{L}31, and show that the field confinement in the disordered \textit{LN} cavities is mainly determined by the local characteristics of the structural disorder as long as the confinement region is far enough from the cavity mirrors and the effective mode localization length is much smaller than the cavity length; under this regime,  the disordered cavity system becomes insensitive to changes in the cavity boundaries and a good agreement with the intensity fluctuation criterion is found for localization. Surprisingly, we find that the Anderson-like localized modes do not appear as new disorder-induced resonances in the main spectral region of the \textit{LN} cavity modes, and, moreover, the disordered DOS enhancement is largest for the disordered waveguide system with the same length. These results are fundamentally interesting for applications such as lasing and cavity-QED, and provide new insights into the role of the boundary condition (e.g., open versus mirrors) on  finite-size slow-light waveguides. They also point out the clear failure of using  models based on the cavity boundaries/mirrors and a single slow-light Bloch mode to describe cavity systems with large \textit{N}, which has been common practise.
\end{abstract}



For more than a decade, unavoidable imperfections arising in the fabrication process of photonic crystal slabs (PCSs) have been shown to have a major influence on the optical performance of PCS-based structures, and an accurate consideration of their effects on the optical properties of PCS devices has been the focus of intense research. Most of these studies have focused on smaller PCS cavities \cite{Gerace2,Noda,Ramunno2,Minkov} and extended  PCS waveguides, in both linear \cite{Gerace,Ramunno,Patterson2} and non-linear media \cite{Krauss,Rossi,Rossi2,Nishan2}, where such imperfections are usually understood as a small amount of intrinsic disorder \cite{Maksim}. Disorder in PCS waveguides is known to be quite rich, and has the capability of inducing phase transitions from extended to localized states \cite{Segev}; in addition, multiple coherent back-scattering phenomena leads to the spontaneous formation of random cavities and the system enters into a simple 1D-like Anderson localization regime \cite{Anderson,John,Patterson}. Such phenomena are intensified near to the band-edges of the guided modes, i.e., in the slow-light regime, where the back-scattering losses increases  approximately as $1/v_g^2$ (with $v_g$ the group velocity) \cite{Hughes}. Interestingly, these unavoidable problems created by random imperfections in PCS waveguides have recently motivated novel approaches in which disorder is required in order to the system works into the desired regime; for instance, recent and important examples are the enhancement of photonic transport in disordered superlattices \cite{Hsieh}, random formation of high-$Q$  cavity modes \cite{Topolancik1,Topolancik2} (where $Q$ is the quality factor), disordered light-matter interaction with applications in cavity-QED \cite{Luca2,Garcia2}, and open transmission channels in strongly scattering media to control light-matter interactions \cite{Sebbah,Riboli,Gurioli}. On the other hand, effects of disorder-induced scattering on PCS cavities have shown to affect mostly the optical quality factor of the cavity modes and induce small fluctuations (in comparison to the $Q$ fluctuations) on their resonant frequencies \cite{Minkov}. These aforementioned studies, in the case of cavities, have been mainly limited to small cavities (such as the $L3$) whose effective sizes are no more than a few periods of the underlying photonic lattice, leading to cavity resonances which fall far from the \textit{W}1 waveguide (a complete row of missing holes) band-edge, i.e., the corresponding system without the cavity mirrors. Nevertheless, recent experiments in slow-light PC lasers show clear evidence that the effects of disorder on large \textit{LN} PCS cavities (with \textit{N} the number of missing holes along the waveguide lattice) are not trivial and poorly misunderstood \cite{Mork}; the lower-frequency cavity modes, which contribute dominantly to the lasing phenomenon, have resonances that approach  the corresponding waveguide band-edge for increasing cavity length, leading to a more sensitive response of those modes to unintentional structural imperfections. Additional localization effects, connected with ideas of Sajeev John's work on localization in light-scattering media \cite{John}, are consequently expected, however, the role of the Anderson phenomenon in such large cavities is not well understood yet, since the non-disordered mode is bounded by the cavity mirrors (the end facet regions of the waveguide); therefore, this is not a truly extended mode that could become a localized state when disorder is introduced in the system. 

\begin{figure}[t!]
\centering
\includegraphics[width=0.45\textwidth]{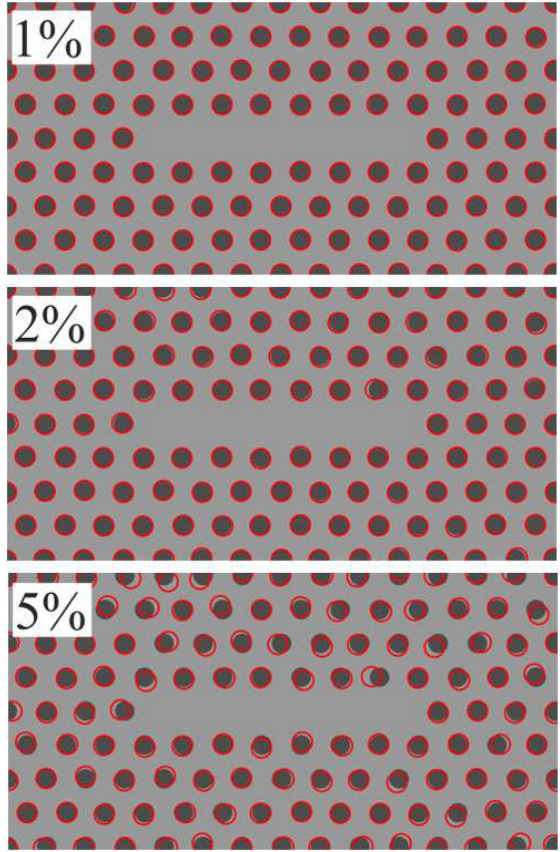}
\caption{Top view of a disordered \textit{L}7 cavity for disorder magnitudes of 1\%, 2\% and 5\% of the PC lattice parameter. The non-disorder (ideal) profile is represented by filled dark-gray circles, while the disordered one is represented by transparent red circles.}\label{fig:disorderp}
\end{figure}

In the present paper, by making  a direct comparison with the disordered waveguide system (namely, without the cavity mirrors), we present an intuitive and detailed explanation of how to approach the Anderson-like localization phenomenon in these longer-length \textit{LN} PCS cavities, which, to the best of our knowledge, has not been addressed in previous works.
For our full vectorial 3D calculations, we use the Bloch mode expansion method (BME) \cite{Savona} and a photon Green function formalism \cite{Vasco} to study the disordered \textit{LN} PCS cavities; in particular, we consider random fluctuations of the in-plane hole positions, with Gaussian probability, as the main disorder contribution to the system, and the corresponding standard variation $\sigma$ of the Gaussian distribution as the main disorder parameter. We show in Figure~\ref{fig:disorderp} disorder realizations of the \textit{L}7 cavity for $\sigma=0.01a$, $\sigma=0.02a$ and $\sigma=0.05a$, i.e., 1\%, 2\% and 5\% of the PC lattice parameter, respectively. The non-disordered (ideal) holes are represented by filled dark-gray circles, while the disordered ones are represented by transparent red circles. By considering $\sigma=0.01a$, i.e., a typical value between typical amounts of intrinsic and deliberate disorder, we compute the optical density of states (DOS) to characterize the spectral properties of the ordered and disordered cavities. We find that additional resonances, namely, different from the ones computed in the non-disordered system, do not appear due to disorder in the cavity mode spectral region, and the corresponding DOS enhancement in the waveguide system (i.e., with no cavity mirrors) is found to be similar to the ones of the largest \textit{LN} cavities and larger than the ones calculated for the small cavity lengths. We also show that the cavity mode $Q$ factors naturally saturate at a specific cavity length rather than become maximized as previously suggested in Ref.~\citenum{Mork}, and the saturation value is bounded by the Anderson modes of the corresponding disordered \textit{W}1 waveguide. In addition, we identify an equivalent behavior for the effective mode volumes, which leads to a reduction of the localization length of the cavity modes; therefore, we find that in general two important effects must be taken into account in disordered \textit{LN} cavities: disorder-induced losses and disorder-induced localization. By means of the intensity fluctuation criterion, commonly employed in disordered photonics \cite{Genack,Luca2,Garcia3,Garcia2}, we assess the Anderson localization phenomenon in the disordered slab cavities and observe Anderson-like localization for cavity lengths equal or larger than \textit{L}31. Moreover, in order to understand the role of the cavity boundary conditions on this phenomenon, we also systematically study the properties of the cavity's fundamental mode for different cavity lengths (ranging from \textit{L}3 to \textit{L}35) and compare with the fundamental disorder-induced mode of the disordered \textit{W}1 waveguide system (which has completely different boundary conditions). 

We also show that, in presence of disorder and under certain conditions, the field confinement inside the cavity region is mainly determined by the local characteristics of the structural disorder and the cavity mirrors do not play an important role in the disordered photonic mode; under this regime, the light confinement displays similar behavior to the ones seen in disordered waveguides, thus suggesting Anderson-like localization in disordered cavities with large \textit{N}. Importantly, since the modal properties become insensitive to the cavity boundary condition, this clearly leads to the breakdown of the single slow-light mode approximation, previously employed to study PCS cavities \cite{Mork,Lalanne}, in the disordered long-cavity regime.

The rest of our paper is organized as follows. In Sec.~\ref{BME} we briefly review the BME method and establish the main parameters of the system. In Sec.~\ref{NondisLn}, results for the cavity resonances, optical quality factor $Q$, effective mode volume $V$ and DOS are presented for non-disordered \textit{LN} PCS cavities. In Sec.~\ref{disLn}, we introduce the model of disorder, compute the disordered DOS, the variance of the normalized intensity distribution, as well as the disordered cavity mode $Q$ factors and the corresponding $V$. Finally, in Sec.~\ref{Rboundary}, we study the role of the boundary condition in the mode localization phenomena and compare with results of Sec.~\ref{disLn}. The main conclusions of the work are presented in Sec.~\ref{Concl}.

\section{Theoretical method}\label{BME}
Large size disordered PCS can be efficiently described by the Bloch Mode Expansion method (BME) under the low loss regime \cite{Savona,Vasco}. To use the BME approach, the disordered magnetic field, $\mathbf{H}_{\beta}(\mathbf{r})$, is expanded in the magnetic field Bloch modes of the non-disordered (ideal) structure, $\mathbf{H}_{\mathbf{k}n}(\mathbf{r})$, with expansion coefficients $U_{\beta}(\mathbf{k},n)$:
\begin{equation}\label{bmeexpansion}
\mathbf{H}_{\beta}(\mathbf{r})=\sum_{\mathbf{k},n}U_{\beta}(\mathbf{k},n)\mathbf{H}_{\mathbf{k}n}(\mathbf{r}),
\end{equation}
where $\mathbf{k}$ and $n$ are the wave vector and band index, respectively, in the first Brillouin zone of the non-disordered structure. Assuming linear, isotropic, non-magnetic, transparent (lossless), and non-dispersive materials, the time-independent Maxwell equations take the form of the following eigenvalue problem when the expansion of Eq.~(\ref{bmeexpansion}) is considered:
\begin{equation}\label{eigenBME}
\sum_{\mathbf{k},n}\left[V_{\mathbf{k}n,\mathbf{k}'n'}+\frac{\omega^2_{\mathbf{k}n}}{c^2}\delta_{\mathbf{k}\mathbf{k}',nn'}\right]U_{\beta}(\mathbf{k},n)=\frac{\omega^2_{\beta}}{c^2}U_{\beta}(\mathbf{k'},n'),
\end{equation}
with disordered matrix elements
\begin{equation}\label{matrixV}
V_{\mathbf{k}n,\mathbf{k}'n'}=\int_{\rm s. cell}\eta(\mathbf{r})\left[\nabla\times\mathbf{H}_{\mathbf{k}n}(\mathbf{r})\right]\cdot\left[\nabla\times\mathbf{H}^{\ast}_{\mathbf{k}'n'}(\mathbf{r})
\right]d\mathbf{r}.
\end{equation}
The integral of Eq.~(\ref{matrixV}) is computed in the {\it total supercell} of the disordered photonic structure and $\eta(\mathbf{r})$ is defined as the difference between the disordered and the non-disordered profiles:
\begin{equation}\label{etadef}
\eta(\mathbf{r})=\frac{1}{\epsilon'(\mathbf{r})}-\frac{1}{\epsilon(\mathbf{r})}.
\end{equation}
The matrix elements of Eq.~(\ref{matrixV}) are conveniently computed from the guided mode expansion (GME) approximation \cite{Lucio}, where the Bloch modes of the non-disordered structure are expanded in the eigenmodes of the effective homogeneous slab. In addition, we employ the so-called photonic ``golden rule'' \cite{Sakoda} to estimate the out-of-plane losses, in which the transition probability $\Gamma_\beta$ from a disordered mode $|\mathbf{H}_{\beta}\rangle$ to a radiative mode $|\mathbf{H}_{\rm rad}\rangle$  (above the light line of the slab) is computed and weighted with the radiative density of states $\rho_{\rm rad}$, so that
\begin{equation}\label{photonicrule}
 \Gamma_\beta = \pi\sum_{\rm rad} \left|\langle\mathbf{H}_{\rm rad}|\hat{\Theta}'|\mathbf{H}_{\beta}\rangle\right|^2\rho_{\rm rad},
\end{equation}
where
\begin{equation}
\hat{\Theta}'=\nabla\times\frac{1}{\epsilon'(\mathbf{r})}\nabla\times,
\end{equation}
is the Maxwell operator associated to the disordered profile. The imaginary part of the frequency, $\Omega_\beta=\Gamma_\beta/(2\omega_\beta)$, obtained through the golden rule in Eq.~(\ref{photonicrule}), and its real part, $\omega_\beta$, determine the optical quality factor of the cavity mode, $Q_\beta=\omega_\beta/(2\Omega_\beta)$. Once the magnetic field of the system is obtained from Eq.~(\ref{eigenBME}), the electric field of the disordered mode is computed via \begin{equation}\label{efield}
\mathbf{E}_{\beta}(\mathbf{r})=\frac{ic}{\omega_\beta\epsilon'(\mathbf{r})}\nabla\times\mathbf{H}_{\beta}(\mathbf{r}),
\end{equation}
which is then normalized through
\begin{equation}\label{enorm}
\int_{\rm s. cell}\epsilon'(\mathbf{r})|\mathbf{E}_{\beta}(\mathbf{r})|^2d\mathbf{r}=1.
\end{equation}
The effective mode volume can be defined as 
\begin{equation}\label{evol}
V_\beta=\frac{1}{\epsilon'(\mathbf{r}_0)|\mathbf{E}_{\beta}(\mathbf{r}_0)|^2},
\end{equation}
where $\mathbf{r}_0$ is usually taken at the antinode position of the electric field peak. 

\begin{figure}[t!]
\centering
\includegraphics[width=0.45\textwidth]{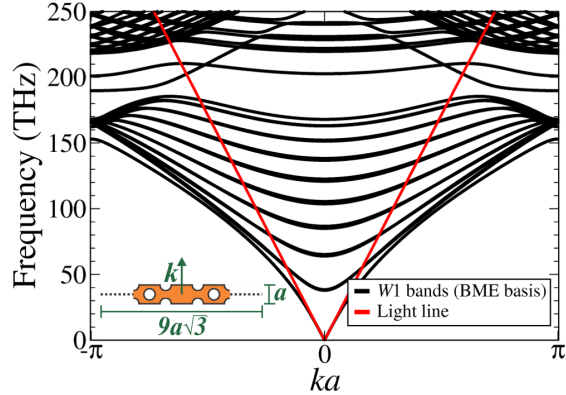}
\caption{ The GME projected band structure of the $a\times 9\sqrt{3}a$ \textit{W}1 PCS waveguide; 200  bands (black curves) are considered in the BME approach.}\label{fig:bands}
\end{figure}

One advantage of the BME method (which is a full 3D\ approach for slabs) is that it is  capable of describing a large volume supercell using the basis of a small supercell or unit cell. Such big supercells can be associated to a system with any degree of disorder, or even to a non-disordered one. In particular, we employ the basis of the \textit{W}1 waveguide, defined in a one-period ($x$ direction) supercell, to describe the \textit{LN} cavity system defined in a supercell of 70 periods ($x$ direction). Both systems have the same size along the $y$ direction which is $9\sqrt{3}a$. We have verified that this supercell of dimensions $70a\times9\sqrt{3}a$ is large enough to achieve the convergence of the cavity modes and to avoid non-physical behavior coming from residual coupling between cavities at neighboring supercells. Using the GME approach, we solve the $a\times9\sqrt{3}a$ \textit{W}1 waveguide, the basis of our \textit{LN} cavity system, by considering the same parameters of Ref.~\citenum{Mork}, i.e., refractive index $n=3.17$ (InP), lattice parameter $a=438$~nm, slab thickness $d=250$~nm and hole radii $r=0.25a$. The projected band structure is shown in Figure~\ref{fig:bands}. Here we have employed 437 plane waves and 1 guided TE mode in the GME basis. Figure~\ref{fig:bands} determines the basis to be used into the BME approach, in particular, we consider $l$ bands and 70 $k$ points, uniformly distributed within the first Brillouin zone, to solve the eigenvalue problem of Eq.~(\ref{eigenBME}). It is well known  that the BME method is quite suitable to optimize large disordered structures \cite{Savona} and provides a fast convergence for the eigenvalues of Eq.~(\ref{eigenBME}), nevertheless, the convergence of the quality factor as a function of the number of bands has been shown to be slow in PCS cavities \cite{Minkov}. We have carried out convergence tests with the non-disordered and disordered cavities and found that $l=200$ is enough to accurately describe the system with a maximum  overall error of 0.05\% and 10\% for the resonant frequencies and quality factors, respectively, with respect to the convergent values.

\section{Non-disordered \textit{LN} cavities}\label{NondisLn}

\begin{figure*}[t!]
\centering
\includegraphics[width=0.9\textwidth]{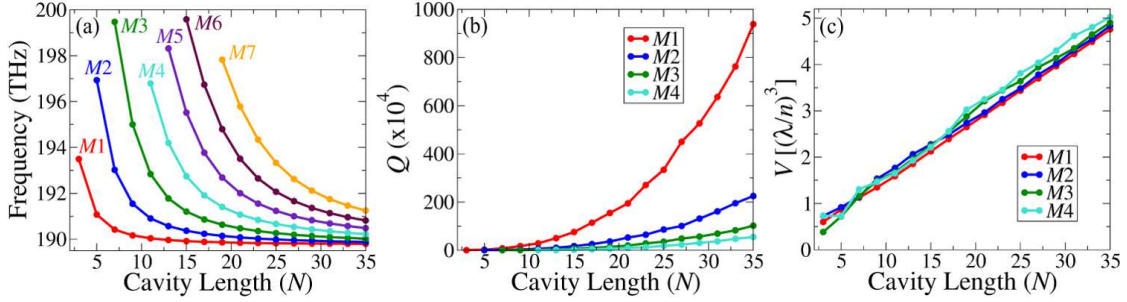}
\caption{(a) Cavity frequencies of the first 7 modes as a function of the cavity length, for the ideal cavities (no disorder). Quality factors (b) and effective mode volumes (c) of the first 4 cavity modes as a function of the cavity length.}\label{fig:sigma0}
\end{figure*}

In order to understand the effects of structural disorder on large \textit{LN} PCS cavities, we first study the non-disordered or ideal lattice system employing the BME approach, where the cavity profile takes the place of the disordered one in Eq.~(\ref{etadef}). Figure~\ref{fig:sigma0} shows the computed results. The cavity resonances as a function of the cavity length are displayed in Figure~\ref{fig:sigma0}a, where one clearly sees a frequency redshift for increasing cavity lengths. Interestingly, the fundamental mode frequency tends to the fundamental \textit{W}1 band-edge, which is predicted at 189.8~THz from our GME calculations; such trend can be understood given the geometry of the system, i.e., large cavities resemble the waveguide system, and consequently, in the limit of $N\rightarrow\infty$ the fundamental cavity resonance has to be exactly 189.8~THz. The fundamental cavity mode is then the one which is expected to be more sensitive to disorder effects for larger cavity lengths; the strong  back-scattering occurring close to the band-edge frequency in the presence of disorder leads to additional localization effects coming from the interplay between order and disorder into the Sajeev John localization phenomenon \cite{John}. These results are also in very good agreement with previous experimental measurements in InP \textit{LN} cavities,  with \textit{N} ranging from 2 to 20 \cite{Mork}; the modal curves $M$1, $M$2, $M$3, $M$4 and $M$5 have exactly the same behavior. We also show, in Figure~\ref{fig:sigma0}b, the optical out-of-plane quality factors, of the first 4 cavity modes, computed with Eq.~(\ref{photonicrule}) as a function of the cavity length. The quality factors of the modes increase for increasing length and those modes whose trend is to fall below the $W$1 light line for very large cavity length are expected to have an unbounded fast increasing of $Q$; and those ones which tend to fall above the light line are expected to have a finite $Q$ in the waveguide limit. Evidently, the fundamental cavity mode, $M$1, displays the largest quality factor (from all the computed modes) given its the trend to approach the fundamental $W$1 band-edge for increasing cavity lengths. These BME $Q$ factors are in very good agreement with recent FDTD calculations in non-disordered \textit{LN} cavities ranging from $L$5 to $L$15 \cite{Cartar}. Correspondingly, we have computed the effective mode volumes of these cavity modes and these are shown in Figure~\ref{fig:sigma0}c. The mode volumes, which are proportional to the localization length of the cavity modes, display an approximately linear increasing for increasing cavity length, recovering the infinite mode volume value (or localization length) in the non-disordered waveguide regime. 

\begin{figure}[t!]
\centering
\includegraphics[width=0.5\textwidth]{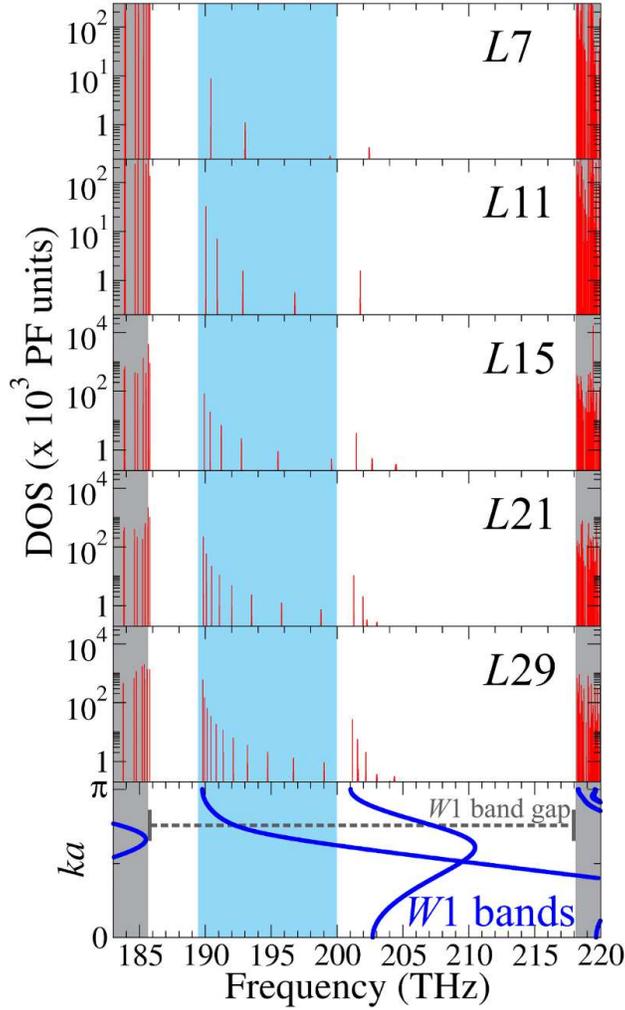}
\caption{The DOS in Purcell factor units (see text) for the non-disordered \textit{L}7, \textit{L}11, \textit{L}15, \textit{L}21, \textit{L}29 cavities. The projected band structure of the non-disordered \textit{W}1 system is shown in the bottom, with the TE-like photonic band gap enclosed by the shaded gray regions, and the frequency region of interest (189.5 to 200 THz), where the main cavity mode appear, has been highlighted in the light-blue shaded area.}\label{fig:DOS0}
\end{figure}

With the aim of studying in more detail the spectral properties of the \textit{LN} cavities we also analyze the DOS of the system, which can be computed using the photonic Green function formalism. The transverse Green's function of the system is expanded in the basis of photonic normal modes \cite{Yao}
\begin{equation}\label{Gfunphc}
\overleftrightarrow{\mathbf{G}}(\mathbf{r},\mathbf{r}',\omega)\approx \sum_\beta \frac{\omega^2\mathbf{E}_\beta^\ast(\mathbf{r}')\mathbf{E}_\beta(\mathbf{r})}{\tilde{\omega}_\beta^2-\omega^2},
\end{equation}
where $\tilde{\omega}_\beta=\omega_\beta-i\Omega_\beta$ is the complex frequency of the mode $\beta$ and the electric field is subject to the normalization condition of Eq.~(\ref{enorm}). Since we are dealing with high $Q$ resonators, in the present formulation we neglect the quasinormal mode aspects of the cavity modes, therefore, the Green's function of Eq.~(\ref{Gfunphc}) and the mode volume definition in Eq.~(\ref{evol}) converge and are expected to be an excellent approximation \cite{Kristensen,Philip}. The local DOS of the system is defined as
\begin{equation}\label{LDOS}
 \rho(\mathbf{r},\omega)=\frac{6}{\pi \omega}\mbox{Im}\left\{\mbox{Tr}\left[\overleftrightarrow{\mathbf{G}}(\mathbf{r},\mathbf{r},\omega)\right]\right\},
\end{equation} 
where the operation Tr[ ] represents the trace over the three orthogonal spatial directions. By integrating Eq.~(\ref{LDOS}) over all space (total DOS), it can be shown that, in units of Purcell factor (which gives the enhancement factor of a dipole emitter), i.e., over the bulk DOS $\omega\sqrt{\epsilon}/(\pi^2c^3)$, the DOS of the system can be written as \cite{Vasco}:
\begin{equation}\label{DOS}
\mbox{DOS}(\omega)=\frac{6\pi c^3}{\omega\epsilon^{3/2}}\mbox{Im}\left\{\sum_\beta\frac{1}{\tilde{\omega}_\beta^2-\omega^2}\right\},
\end{equation}
where $\epsilon$ is the bulk/slab dielectric constant (associated to InP in our case).  Figure~\ref{fig:DOS0} shows the computed DOS in a log-scale for the $L$7, $L$11, $L$15, $L$21 and $L$29 cavities, and the band structure of the $W$1 is shown in bottom where the shaded gray regions enclose the TE-like photonic band gap of the $W$1 PC. The frequency region of interest, 189.5 to 200 THz [same of Figure~\ref{fig:sigma0}(a)], is shaded in light-blue. As expected from results of Figure~\ref{fig:sigma0}, the DOS intensity increases, with deceasing peak-width, for increasing cavity length at the cavity mode frequencies (at resonance with the cavity mode $\beta$ the DOS is proportional to $Q_\beta$), and the frequency spacing between the peaks decreases as \textit{N} increases. Since smaller \textit{LN} cavities lead to smaller average refractive index in the PC slab (more holes), the photonic band-edges of the cavity structures, enclosing the photonic band gap, are slightly blue-shifted (with respect to the $W$1 case) as the cavity length decreases. In the shaded blue region, we found a total of 3, 4, 6, 8 and 11 cavity resonances of the $L$7, $L$11, $L$15, $L$21 and $L$29 cavities, respectively.

\section{Disordered \textit{LN} cavities}\label{disLn}

Planar PCS are always subject to a small degree of unintentional imperfections, coming from the fabrication process, which can be understood as  ``intrinsic  disorder'' in the dielectric lattice. The problem of modelling disorder in PCS has been addressed in the literature by considering simple hole-size or hole-position random fluctuations \cite{Nishan,Gerace}, as well as more sophisticated models where a random-correlated surface roughness is introduced in the holes surfaces\cite{Minkov3,Nishan3}. Here, we consider random fluctuations of the hole positions as the main disorder contribution. Such a model of disorder has shown to be accurate  for understanding the effects of unintentional structural disorder in PCS waveguides \cite{Garcia}, and  is similar to random fluctuations in the hole sizes \cite{Vasco}. 

Starting from the non-fluctuated position $(x^{(0)}_m,y^{(0)}_m)$ of the $m$-th hole, we consider random fluctuations $\delta$ with Gaussian probability within the BME supercell:
\begin{equation}\label{fluctuations}
(x_m,y_m)=(x^{(0)}_m+\delta x,y^{(0)}_m+\delta y),
\end{equation}
where $(x_m,y_m)$ is the fluctuated position. We adopt the standard deviation of the Gaussian probability distribution $\sigma=\sigma_x=\sigma_y$ as our disorder parameter. In the present work, we set a medium quantity of structural disorder $\sigma=0.01a$, i.e., 1\% of the lattice parameter, corresponding to a disorder magnitude which is between typical amounts of intrinsic and deliberate disorder values that have been used previously  \cite{Nishan}. In all the disordered cases that will be shown below, we have considered 20 independent statistical realizations of the disordered system, and we have verified that this number of instances is enough to describe the main physics of the disordered PCS structure. 

\begin{figure*}[t!]
\centering
\includegraphics[width=1.0\textwidth]{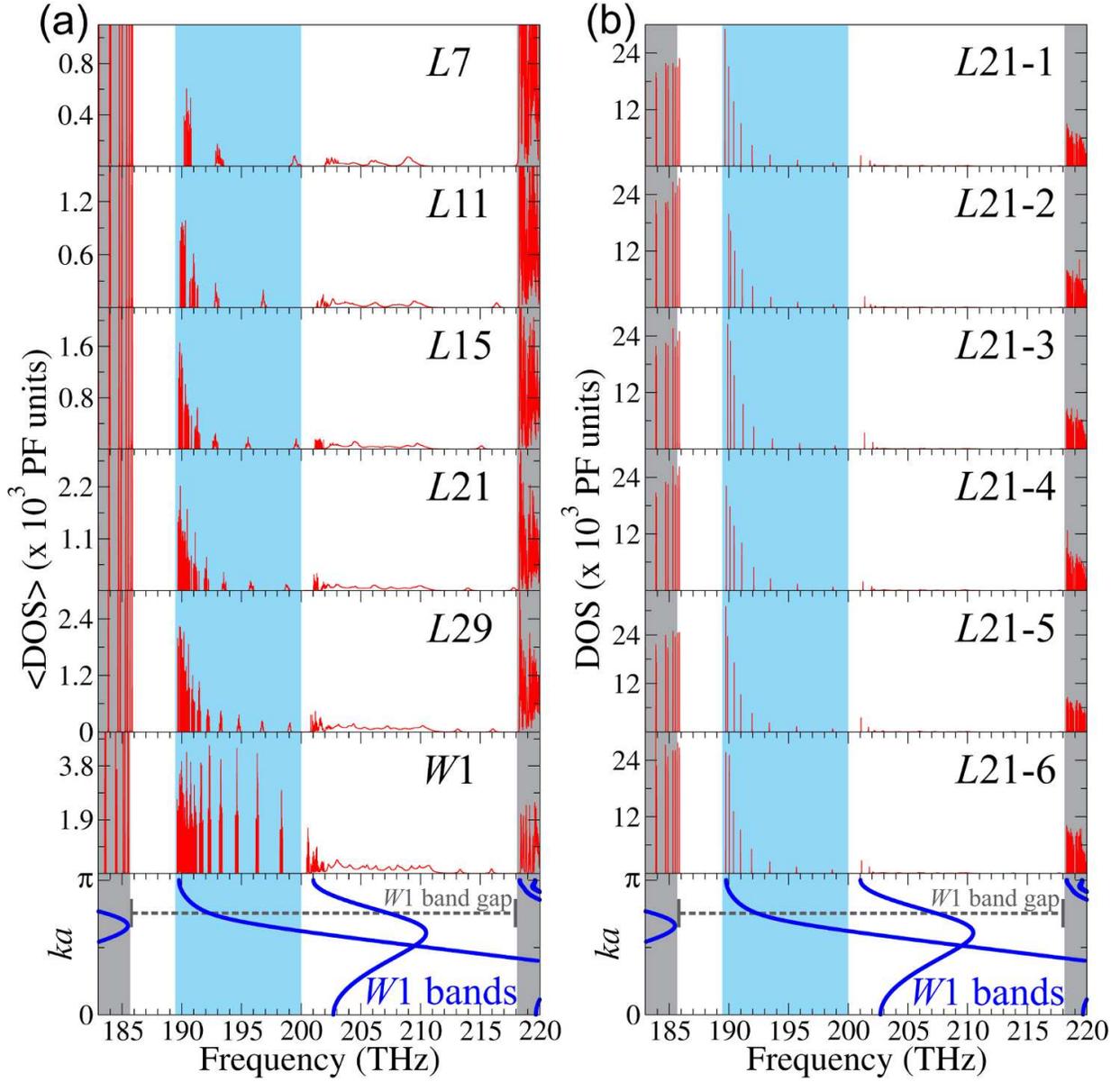}
\caption{(a) Averaged DOS in Purcell factor units (see text) for the \textit{L}7, \textit{L}11, \textit{L}15, \textit{L}21, \textit{L}29 cavities and the corresponding \textit{W}1 waveguide. We have considered 20 independent statistical instances  and the amount of intrinsic disorder is $\sigma=0.01a$. (b) The DOS, in units of Purcell factor, is shown for six independent statistical realizations of the disordered \textit{L}21 cavity with $\sigma=0.01a$. The projected band structure of the non-disordered \textit{W}1 system is shown in bottom of both (a) and (b), the frequency region of interest (cavity modes) has been highlighted in the light-blue shaded area  and the regions enclosing the band gap of the non-disordered waveguide are shaded in gray.}\label{fig:DOS}
\end{figure*}

\begin{figure}[t!]
\centering
\includegraphics[width=0.45\textwidth]{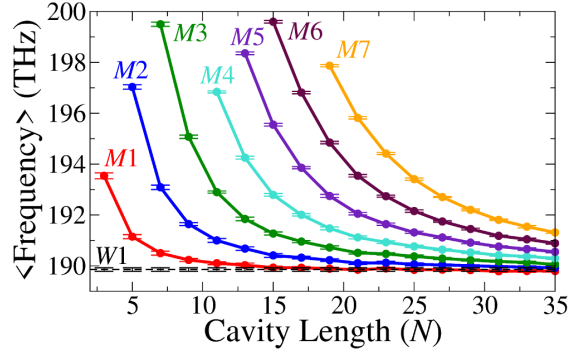}
\caption{Average frequencies of the first 7 cavity modes as a function of the cavity length.  The horizontal dashed line corresponds to the Anderson modes of the disordered \textit{W}1 waveguide (ensemble average). The standard deviation is represented by the error bars. We have used 20 independent statistical instances with an amount of intrinsic disorder of $\sigma=0.01a$.}\label{fig:modes01}
\end{figure}

In~\ref{fig:DOS}a, we show the averaged DOS over 20 statistical realizations of the disordered system for the \textit{L}7, \textit{L}11, \textit{L}15, \textit{L}21, \textit{L}29 cavities and the \textit{W}1 system. The projected band structure of the non-disordered \textit{W}1 waveguide is also shown inside the reduced Brillouin zone and, as in Figure~\ref{fig:DOS0}, the shaded gray regions enclose the TE-like photonic band gap of the waveguide. The $\braket{\mbox{DOS}}$ displays sets of sharp resonances inside the photonic band gap of the system corresponding to the cavity modes; the number of resonances increases as the cavity length increases recovering the Lifshitz tail in the waveguide limit, at both fundamental and second waveguide band-edges. Interestingly, the Purcell enhancement determined by the discrete resonances in the \textit{W}1 waveguide, inside the shaded region (dominant cavity mode spectral region), is similar to the \textit{L}29 case and larger than those of the smaller \textit{LN} cases, suggesting that the random cavity modes, spontaneously induced by disorder in the waveguide, are as good or even better than designed/engineered cavity modes of the perfect lattice. From Figure~\ref{fig:DOS}a, it is possible to picture the overall trend of the system but it is difficult to conclude if new modes are appearing in the sets of sharp resonances due to disorder, which would be, at first insight, associated to Anderson-like localized modes inside the cavities. To identify  if such new modes are in fact appearing inside the cavities, we show in Figure~\ref{fig:DOS}b six independent statistical instances of DOS for the \textit{L}21 cavity. Here, individual sharp peaks, corresponding to the system resonances, can easily be resolved. From this figure, it is possible to see that the number of peaks in the shaded region is the same of the corresponding non-disordered DOS of Figure~\ref{fig:DOS0}, i.e, 8 resonant peaks, and it is always the same for all disordered realizations of the cavity. We have also seen the same behavior for all disordered instances of the cavities studied in this work. Therefore, we can conclude that no new modes  appear in the \textit{LN} cavities due to disorder (at least not in the cavity modes' spectral region).  This result is also valid for larger amounts of disorder as long as the cavity resonances are not mixed with the induced Anderson modes at the band edges of the PC band gap. As in the case of the non-disordered systems, the band-edges enclosing the photonic band gap of the disordered cavities, are also slightly blue-shifted for decreasing cavity length. Taking into account the fact that no new resonances are induced by disorder in the light-blue shaded area of Figure~\ref{fig:DOS}(b), we have computed the averaged frequencies of the cavity modes as a function of the cavity length and the results are shown in Figure~\ref{fig:modes01}, where  the standard deviation is represented by the error bars. The averaged frequency spectrum of the disordered system is slightly blue-shifted, in comparison to Figure~\ref{fig:sigma0}a, but all the functional features are equivalent to the non-disordered case. In particular, we note larger fluctuations for smaller cavities, which is in agreement with the approximate scaling of the disorder-induced resonance shifts with the inverse mode volume \cite{Ramunno2}. These results show then that the cavity frequencies, apart from small fluctuations around an average value, are not fundamentally affected when sizable amounts of disorder are present in the \textit{LN} cavity system.

\begin{figure*}[t!]
\centering
\includegraphics[width=1.0\textwidth]{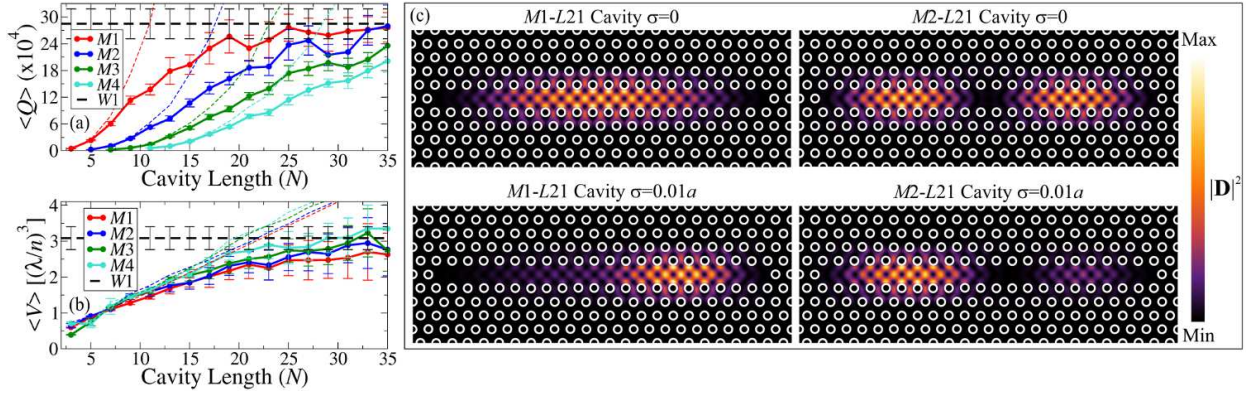}
\caption{Averaged (a) quality factor and (b) mode volumes as a function of the cavity length.  The standard deviation is represented by the error bars and the thin dashed lines correspond to the results without considering disorder. We have considered 20 independent statistical instances  and the amount of intrinsic disorder is $\sigma=0.01a$. The horizontal dashed lines are for the Anderson modes of the disordered \textit{W}1 waveguide (ensemble average)  with the error bars representing the corresponding standard deviation. (c) Intensity profiles in the center of the slab $|\mathbf{D}(x,y,z=0)|^2$ of the first two cavity modes for the \textit{L}21 cavity; we have used  $\sigma=0.0$ and $\sigma=0.01a$.  The non-disordered (ideal) and disordered PC profile are correspondingly represented by the thick white circles.}\label{fig:QV01}
\end{figure*}

Different from the cavity resonance properties, the quality factors and effective mode volumes display much more dramatic behaviors in the presence of disorder. Figure~\ref{fig:QV01}a shows the averaged $Q$ as a function of the cavity length for the first 4 cavity modes, and the averaged $Q$ values of the Anderson modes in the \textit{W}1 system are represented by the horizontal dashed line.  The associated standard deviations are represented by the error bars and the corresponding non-disorder results (\ref{fig:sigma0}) are shown in thin dashed lines. When disorder is considered, the quality factors of the cavity modes are reduced (as expected), and those modes which are closer to the band-edge of the waveguide, i.e., in the slow-light regime, are much more sensitive to disorder effects, as is the case of the fundamental cavity mode \textit{M}1 (red curve), which decreases from $\sim938\times10^4$ (see Figure~\ref{fig:sigma0}b) to $\sim27\times10^4$ for the \textit{L}35 cavity, and the second mode \textit{M}2 (blue curve)  decreases from $\sim225\times10^4$ (see Figure~\ref{fig:sigma0}b) to $\sim27\times10^4$ for the same cavity. By comparing Figure~\ref{fig:QV01}a with Figure~\ref{fig:sigma0}b (or with the thin dashed lines included in Figure~\ref{fig:QV01}a) we see that such a reduction is less strong for those modes which are not close to the waveguide band-edge, i.e., far from the slow-light regime. Moreover, we have not identified any specific cavity length at which the quality factor is maximized when disorder is considered, which was previously suggested in Ref.~\citenum{Mork} as a possible explanation of the threshold minimization in slow-light PC lasers (when the cavity length was increased). In contrast, we have found that the quality factor increases for increasing cavity length and eventually saturates, with small oscillations, below the averaged $Q$ of the Anderson modes of the corresponding waveguide system. Since the mean quality factor of such Anderson modes depends on the degree of disorder, the cavity length at which the mean $Q$ saturates also depends on the specific value of $\sigma$, implying that the $Q$ performance of the cavities is roughly bounded by the average $Q$ performance of the \textit{W}1 Anderson modes. The averaged mode volumes are shown in Figure~\ref{fig:QV01}b, where it is demonstrated that $V$ is reduced in large cavities, with respect to the non-disordered case (see thin dashed lines or~\ref{fig:sigma0}c), in the presence of disorder, and, as for the case of $Q$, the volume of the cavities also remains bounded by the average $V$ of the Anderson-like modes in the \textit{W}1 waveguide. The disorder-induced localization suggested by Figure~\ref{fig:QV01}b is clearly seen in Figure~\ref{fig:QV01}c, where the intensity profile $|\mathbf{D}(\mathbf{r})|^2$ [with $\mathbf{D}(\mathbf{r})=\epsilon_0\epsilon(\mathbf{r})\mathbf{E}(\mathbf{r})$] is shown for the first two cavity modes with $\sigma=0$ and $\sigma=0.01a$ in the upper and bottom panels, respectively. The non-disordered and disordered profiles are correspondingly represented by thick white circles. In Figure~\ref{fig:QV01}c, the field is concentrated in a smaller region within the cavity, implying a reduction of the effective localization length of the mode. Thus, as remarked earlier, we conclude that structural disorder in \textit{LN} cavities causes disorder-induced losses and \textit{disorder-induced localization}.  

\begin{figure}[t!]
\centering
\includegraphics[width=0.45\textwidth]{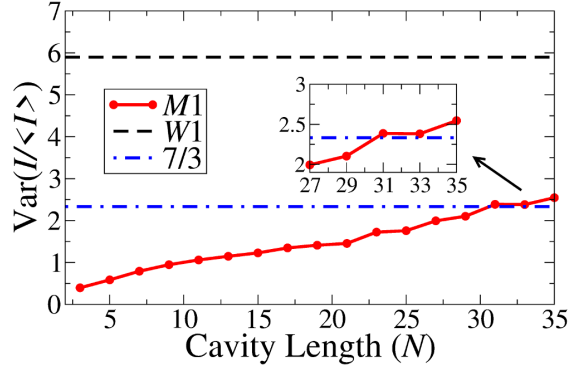}
\caption{Variance of the normalized intensity distribution (the local DOS has been averaged in the frequency region of the fundamental cavity mode) as a function of the cavity length (red line). The corresponding value for the $W$1 waveguide is 5.9 (horizontal black dashed line) and the critical value 7/3 is represented by the horizontal blue chain line.}\label{fig:VarInt}
\end{figure}

 In order to assess the quasi-1D Anderson localization phenomenon in the disordered \textit{LN} cavity system, a clear broadening of the intensity distribution has to be seen with respect to the Rayleigh distribution \cite{Genack}. Given the proportionality between the vertical emitted intensity with the local DOS through the out-of-plane radiative decay \cite{Nishan,Garcia2}, we employ the local DOS fluctuations to obtain the intensity fluctuations, i.e., $\mbox{Var}\left(I/\braket{I}\right)=\mbox{Var}\left[\rho(\mathbf{r},\omega)/\braket{\rho(\mathbf{r},\omega)}\right]$  \cite{Shapiro,Carminati}. The variance of the normalized intensity distribution, $\mbox{Var}(I/\braket{I})$, is then computed by employing the local DOS of all disordered realizations along the cavity direction, $\rho(x,y=0,z=0,\omega)$, averaged over the desired frequency range \cite{Garcia3}. We show in Figure~\ref{fig:VarInt}, the variance of the intensity fluctuations in the averaged frequency region of the fundamental cavity mode (usually the most important for practical applications) and the corresponding value obtained for the disordered $W$1 system, 5.9, which is in good agreement with previous experimental measurements in similar disordered $W$1 waveguides \cite{Luca2}. By assuming that the intensity fluctuations follows the same statistics of the transmission fluctuations, the system falls into the Anderson localization regime if $\mbox{Var}(I/\braket{I})$ exceeds the critical value 7/3 \cite{Genack,Luca2}, i.e., the blue chain-check line in~\ref{fig:VarInt}. Consequently, we observe Anderson-like characteristics for cavity lengths equal or larger than $L$31 (see inset of the figure). 

\begin{figure*}[t!]
\centering
\includegraphics[width=1.0\textwidth]{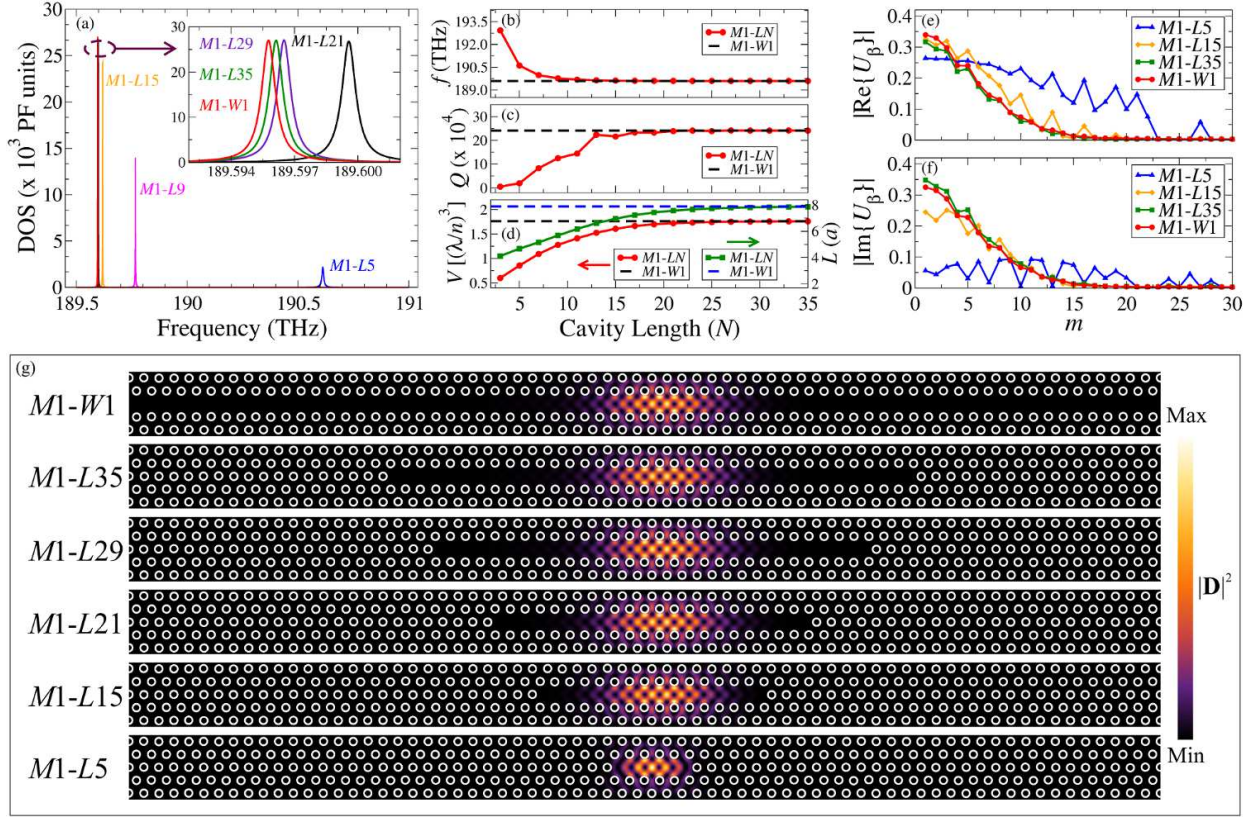}
\caption{(a) Computed DOS of the fundamental cavity mode \textit{M}1 for several cavity lengths by considering \textit{exactly} the same disorder realization with $\sigma=0.01a$. The fundamental disorder-induced mode of the disordered \textit{W}1 waveguide, \textit{M}1-\textit{W}1, is also shown. (b) Frequency of the fundamental cavity mode as a function of the cavity length; as in (a), \textit{exactly} the same disorder realization is considered with $\sigma=0.01a$. The frequency of the  \textit{M}1-\textit{W}1 mode is represented by the horizontal dashed line. (c) Quality factors of the modes shown in panel (b). (d) Effective mode volumes (left) and localization length (right) of the modes shown in panel (b). Absolute value of the (e) real and (f) imaginary part of the expansion coefficient $U_{\beta}(m)$ for $\beta=$\textit{M}1, in the \textit{L}5, \textit{L}15 and \textit{L}35 cavities, and $\beta=$\textit{M}1-\textit{W}1, in the disordered \textit{W}1 waveguide, where $m=(\mathbf{k},n)$ is a global index such that $U_{M{\rm{1-}}W\rm{1}}(m)>U_{M{\rm{1-}}W\rm{1}}(m+1)$. (g) Intensity profiles in the center of the slab $|\mathbf{D}(x,y,z=0)|^2$ of the fundamental mode in the corresponding \textit{W}1 and cavity systems. The disordered profile is represented by thick white circles.}\label{fig:boundary}
\end{figure*}

\section{Role of the cavity boundary conditions}\label{Rboundary}

\begin{figure}[t!]
\centering
\includegraphics[width=0.45\textwidth]{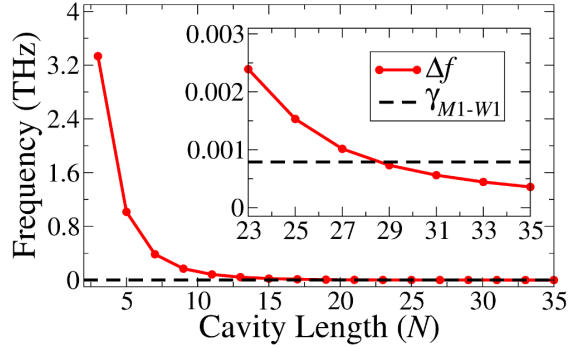}
\caption{Frequency difference $\Delta f=f_{M\rm{1}}-f_{M{\rm{1-}}W\rm{1}}$ as a function of the cavity length. The photonic radiative linewidth of the mode \textit{M}1-\textit{W}1 is represented by the horizontal dashed line.}\label{fig:dfA}
\end{figure}

The apparent saturation of $Q$ and $V$ in Figure~\ref{fig:QV01}a and Figure~\ref{fig:QV01}b suggests that when disorder is considered, on average, there is a minimum cavity length at which the boundary cavity condition is not fundamentally relevant to the cavity mode. In order to address the role of the cavity boundaries on the mode confinement, we now systematically study the DOS and the field profiles for several cavity lengths using \textit{exactly} the same disorder realization, i.e., the same distribution of disordered holes. Results of this analysis are shown in Figure~\ref{fig:boundary}. The DOS associated to the fundamental mode is displayed in Figure~\ref{fig:boundary}a for the \textit{L}5, \textit{L}15, \textit{L}21, \textit{L}29, \textit{L}35 cavities and the fundamental disorder-induced mode of the disordered \textit{W}1 waveguide, i.e., the lower frequency mode appearing in the diagonalization within the TE-like band gap of the \textit{W}1 PC slab, which we have denoted as \textit{M}1-\textit{W}1. The frequency difference between the fundamental modes of the different cavities decreases for increasing cavity length (as expected from Figure~\ref{fig:modes01}) and, as already seen in the averaged DOS of Figure~\ref{fig:DOS}a, the intensity of the DOS is largest for the mode \textit{M}1-\textit{W}1 and quite similar to the corresponding intensity for cavity lengths larger than \textit{L}21. We also show in Figure~\ref{fig:boundary}b, Figure~\ref{fig:boundary}c and Figure~\ref{fig:boundary}d the frequency, quality factor and effective mode volume (including the localization length), respectively, of the fundamental cavity mode, \textit{M}1, as a function of the cavity length by (again) considering exactly the same disordered PC. The corresponding values of the fundamental disorder-induced mode \textit{M}1-\textit{W}1 are represented by the horizontal dashed lines. The localization length in Figure~\ref{fig:boundary}d is computed using the inverse participation number as described in Refs.~\citenum{Savona} and \citenum{Vasco}. From these results we can clearly see that, for large length disordered cavities, the cavity fundamental mode and the \textit{M}1-\textit{W}1 mode are totally equivalent, and this hypothesis is further strengthened in Figure~\ref{fig:boundary}e and Figure~\ref{fig:boundary}f, where the absolute value of ${\rm Im}\{U_\beta(m)\}$ and ${\rm Re}\{U_\beta(m)\}$, i.e., the expansion coefficient in Eq.~(\ref{bmeexpansion}), are respectively shown for the fundamental mode of the \textit{L}5, \textit{L}15 and \textit{L}35 cavities, $\beta=M1$, and the mode $\beta=$\textit{M}1-\textit{W}1 in the disordered waveguide [the coefficients are organized such that $U_{M{\rm{1-}}W\rm{1}}(m)>U_{M{\rm{1-}}W\rm{1}}(m+1)$ given the global index $m=(\mathbf{k},n)$]; the difference between the expansion coefficients of $\beta=$\textit{M}1-\textit{W}1 and $\beta=$\textit{M}1 modes becomes negligible for large cavities lengths, thus demonstrating the equivalence between these two confined states in the presence of disorder. The similarity between the fundamental cavity modes and the \textit{M}1-\textit{W}1 mode can also be understood from Figure~\ref{fig:boundary}g, where we have plotted their intensity profiles with the disordered profile represented by thick white circles. For strongly localized Anderson modes, which is the case of \textit{M}1-\textit{W}1, the cavity boundary condition does not significantly affect the mode distribution as long as the boundaries of the cavity are far from the localization region, then the confinement in such large cavities can be mainly determined by the specific local characteristics of the structural disorder than by the cavity mirrors. It is important to stress again that the localized mode must be positioned far from the cavity boundaries and its localization length 
must be much smaller than the cavity length in order to establish this equivalence; in 
such a regime, the cavity boundaries do not play an important role and the confinement is mainly determined by Anderson-like localization phenomena. We show in Figure~\ref{fig:dfA} the frequency difference between the fundamental cavity mode and the fundamental disorder-induced mode of the disordered \textit{W}1 waveguide, i.e.,  $\Delta f=f_{M\rm{1}}-f_{M{\rm{1-}}W\rm{1}}$, as a function of the cavity length, compared with the photonic radiative linewidth of the \textit{M}1-\textit{W}1 mode (horizontal dashed line). As clearly seen in the inset of the figure, the frequency difference between the fundamental disordered cavity modes and the Anderson mode of the disordered waveguide is  blurred for cavity lengths equal or larger than $L$31, suggesting that the fundamental mode of these disordered systems is not significantly sensitive to the boundary cavity condition for $N\geq31$. The sensitivity of the system eigenvalues to the boundary conditions has been previously employed as a criterion for Anderson localization in semiconductors \cite{Thouless}; here, we also found a very good agreement with the intensity fluctuations criterion employed in disordered photonics, thus reinforcing our conclusions.

As a consequence of these results, we highlight that models based on the cavity boundaries and single slow-light Bloch modes to describe cavity modes with large \textit{N}, as it is the case of the effective Fabry-P\'erot resonator \cite{Lalanne,Mork}, will be problematic for predicting the behavior of Anderson-like localized modes inside the disordered cavity region. The main reason, as addressed above, is that the confined state cannot in general be understood as two counter-propagating Bloch modes reflecting between the two cavity mirrors, since the confined mode does not necessarily feel the cavity boundaries when the system falls into the Anderson-like localization regime, i.e., when the slow-light Bloch modes are subject to several strong back-reflections originated in the local distribution of disordered holes.

\section{Conclusions}\label{Concl}

We have presented a detailed study of the effects of structural disorder on PCS \textit{LN} cavity modes by employing a fully 3D BME approach and Green function formalism. By considering in-plane $\sigma$ magnitudes, which are between typical amounts of intrinsic and deliberate disorder, we found, on the one hand, that disorder  induces fluctuations in the fundamental resonance frequencies and no new additional modes  appear in the usual cavity spectral region of interest. However, disorder has a strong influence on the quality factor and mode volume of these modes, especially for longer length cavities. We also studied the onset of localization modes for the cavity structures,  over a wide range of cavity lengths and frequencies;
interestingly, our  averaged DOS results suggest that the Anderson-modes of the disordered slow-light waveguides are as good or even better than  non-optimized disordered \textit{LN} cavities in terms of  enhancing the total DOS, which opens new possibilities on designing high quality disordered modes,  e.g., by taking the disordered profile as an improved design, that could surpass the performance of state-of-the-art engineered PCS cavities---which has been recently demonstrated experimentally \cite{Luca}. Furthermore, we have found that the mean mode quality factors and respectively mean effective mode volumes saturate for a specific cavity length scale (which depends on the disorder parameter), and they are bounded by the averages of the Anderson modes in the corresponding \textit{W}1 waveguide system. Next, we have shown that: (i) the quality factor is not maximized at a specific length in disordered \textit{LN} cavities as previously suggested in Ref.~\citenum{Mork};  (ii), apart from disorder-induced losses, disorder-induced localization becomes critically important in longer length disordered \textit{LN} cavities; and (iii), by means of the intensity fluctuation criterion we have observed Anderson-like localization for cavity lengths equal or larger than $L$31. In order to further understand the role of the cavity boundary conditions on the cavity mode confinement, we also systematically studied the effects of the cavity mirrors on the fundamental cavity mode (which is usually the most relevant one for practical applications); we found that as long as the confinement region is far from the cavity boundaries and the effective mode localization length is much smaller than the cavity length, the photonic confinement is mainly determined by Anderson-like localization and the mirrors of the cavity do not play an important role on the definition of the resonant frequency, quality factor and effective volume of the mode in system; this finding is significant as these are the key quantities to understand the light-matter coupling parameters in numerous applications such as lasing, sensing,  nonlinear optics  and cavity-QED. Clearly, for significantly long $LN$ cavities, once can expect a similar performance from either $W1$ waveguides or the cavities, wherein the physics of the cavity models can change significantly.  

\begin{acknowledgement}

This work was supported by the Natural Sciences and Engineering Research Council of Canada (NSERC) and Queen's University, Canada. We gratefully acknowledge Jesper
M\o rk and Luca Sapienza for useful suggestions and discussions. This research was enabled in part by computational support provided by the Centre for Advanced Computing (http://cac.queensu.ca) and Compute Canada (www.computecanada.ca). 

\end{acknowledgement}





\providecommand{\latin}[1]{#1}
\providecommand*\mcitethebibliography{\thebibliography}
\csname @ifundefined\endcsname{endmcitethebibliography}
  {\let\endmcitethebibliography\endthebibliography}{}

\newpage

\end{document}